\documentclass[twocolumn,aps,prb,showpacs,superscriptaddress,amsmath,amssymb]{revtex4}
\usepackage{amssymb}

\usepackage{amsmath}
\usepackage{amssymb,braket}
\usepackage{epsfig}
\usepackage{color}
\usepackage{graphicx}
\usepackage{verbatim}
\usepackage{dcolumn}
\usepackage{bm}

\renewcommand{\v}[1]{{\bf #1}}

\newcommand{\be}{\begin{equation}}
\newcommand{\ee}{\end{equation}}
\newcommand{\nn}{\nonumber \\}

\newcommand{\ba}{\begin{eqnarray}}

\newcommand{\ea}{\end{eqnarray}}

\newcommand{\bw}{\begin{widetext}}
\newcommand{\ew}{\end{widetext}}

\newcommand{\bpm}{\begin{pmatrix}}
\newcommand{\epm}{\end{pmatrix}}

\begin{document}
\title{Landau level states on a topological insulator thin film}

\author{Zhihua Yang}
%\email[Electronic address:$~~$]{zhihuayang@skku.edu}
\affiliation{ Department of Physics and BK21 Physics Research
Division, Sungkyunkwan University, Suwon 440-746, Korea}
\affiliation{ Xinjiang Technical Institute of Physics $\&$
Chemistry, Chinese Academy of Sciences,  Urumqi 830011, China}
\author{Jung Hoon Han}
\email[Electronic address:$~~$]{hanjh@skku.edu} \affiliation{
Department of Physics and BK21 Physics Research Division,
Sungkyunkwan University, Suwon 440-746, Korea}
%\date{\today}

\begin{abstract} We analyze the four-dimensional Hamiltonian
proposed to describe the band structure of the single-Dirac-cone
family of topological insulators in the presence of a uniform
perpendicular magnetic field. Surface Landau level(LL) states
appear, decoupled from the bulk levels and following the quantized
energy dispersion of a purely two-dimensional surface Dirac
Hamiltonian. A small hybridization gap splits the degeneracy of the
central $n=0$ LL with dependence on the film thickness and the field
strength that can be obtained analytically. Explicit calculation of
the spin and charge densities show that surface LL states are
localized within approximately one quintuple layer from the surface
termination. Some new surface-bound LLs are shown to exist at a
higher Landau level index.
\end{abstract}


\maketitle

\section{Introduction}

Insulating materials with topologically protected surface states
known as topological insulators (TIs) are a matter of great current
interest\cite{kane-hasan,qi-zhang,moore}. The surface metallic states
in this new class of materials is characterized by Dirac-like
quasiparticle dispersion, and a one-to-one correspondence between
momentum  and spin quantum numbers of the single-particle states thus
representing an extreme form of spin-orbit coupling. Both these
aspects have been confirmed for the first time in Bi$_x$Sb$_{1-x}$
family\cite{kane-fu} of topological insulators by
ARPES\cite{ARPES-on-BiSb} and STM\cite{STM-on-BiSb} studies.

More recently, a lot of experimental efforts has been given to the
synthesis and characterization of Bi$_2$Se$_3$, Bi$_2$Te$_3$, and
Sb$_2$Te$_3$\cite{ARPES-on-Bi2Te3,STM-on-Bi2Se3,ong,Bi2Se3-exp,
thin-film-TI-exp-Japan,thin-film-TI-exp-China,STM-B-thin-film,hanaguri}
following the prediction of their topological
behavior\cite{Bi2Se3-theory,Bi2Se3-theory2}, due to their simple
surface band structure consisting of a single-cone Dirac spectrum
centered at the $\Gamma$-point and a relatively large band gap.
Topological insulators of the single-Dirac-cone family in the
thin-film form has been synthesized by a number of
groups\cite{thin-film-TI-exp-Japan,thin-film-TI-exp-China,STM-B-thin-film}.
Theoretically, the thin-film TIs bear close analogy to another
heavily studied topological material, i.e.
graphene\cite{castro-neto}. For instance, the well-known pair of
valley-degenerate Dirac bands of graphene becomes the top and bottom
surface Dirac bands of TIs with finite thickness. Perpendicular
magnetic field quantizes the surface Landau levels (LLs) with the
energies that scale with the LL index $n$ as $\pm \sqrt{n}$ in TIs as
well as in graphene.

Previous treatments of the magnetic field effect on TI surface
started from the two-dimensional (2D) Dirac Hamiltonian focusing only
on the surface electronic states and ignoring the bulk states
altogether\cite{qi,SQShen-LL}. These methods relied on first
projecting the bulk Hamiltonian to the surface, obtaining the 2D
Dirac model, then including the field effect by way of Peierls
substitution. In another vein, several recent papers theoretically
examined the properties of a thin slab of TI in which the bulk and
surface electronic states are treated on an equal
footing\cite{linder,liu,shen} in the absence of the magnetic field.
It is thus natural to consider how the magnetic field effect plays
out for a thin film geometry of TI, following the spirit of solving
the bulk Hamiltonian adopted in Refs. \onlinecite{linder,liu,shen}.
In fact, an attempt of precisely this sort has been made in a recent
paper by Liu \textit{et al}.\cite{Bi2Se3-theory2} Here, the authors
solved the $4\times4$ tight-binding Hamiltonian with the Peierls
substitution for the magnetic field and even including the Zeeman
field coupling. We point out in this paper that the method adopted in
Ref. \onlinecite{Bi2Se3-theory2} does not treat the surface and bulk
electronic states simultaneously, and as a result the bands arising
from surface LLs penetrate into the bulk LL states, while physically
such overlapping of energy levels will not occur.

Our approach follows closely the spirit of zero-field case studied in
Refs. \onlinecite{linder,liu,shen} and takes care of the boundary
conditions properly. Some parts of our report are technical, dealing
with the characteristic equation resulting from the boundary
conditions and the methods of solving them. Several physically
meaningful results follow from our analysis. First, hybridization of
the zeroth-LL states localized on the top and the bottom surfaces for
a sufficiently thin sample is shown to manifest itself as the
splitting of the degeneracy of zeroth-LL states with the gap
magnitude that can be calculated analytically. The zeroth-LL gap size
oscillates with the film thickness. Our finding naturally
extrapolates a similar observation of the gap oscillation observed
previously\cite{linder} to finite magnetic field. Interestingly, we
find that a new kind of surface-bound LL states appear for higher-LL
indices where the conventional surface LL band of $\sim \sqrt{n}$
variety has merged into the bulk continuum. Justification of the new
surface LLs is made on the basis of careful numerical study and an
approximate analytic solution of the characteristic equation.
Properties of the bulk single-particle states for higher-LL indices
are examined in detail. Finally, both charge and spin density
profiles of the surface LLs at low LL indices along the thickness of
the sample are explicitly worked out.

In Sec. \ref{sec:formulation} we formulate the LL problem based on
the $4\times4$ Hamiltonian proposed previously for
Bi$_2$Se$_3$-family of topological insulators. Boundary conditions
are imposed on the two surface layers for a thin-film geometry and
characteristic equations are derived in Sec. \ref{sec:LL}. In Sec.
\ref{sec:results} several physical results are shown and its
relevance to recent STM are discussed. Summary of results and an
outlook is given in Sec. \ref{sec:conclusion}. Technical discussion
for the new surface-bound LLs can be found in the Appendix.

\section{Formulation}
\label{sec:formulation}

The 3D tight-binding Hamiltonian proposed as a minimal model for
single-Dirac-cone family of TIs first in Ref.
\onlinecite{Bi2Se3-theory} and detailed in Ref.
\onlinecite{Bi2Se3-theory2} is

\ba H(\v p) = \varepsilon(\v p) \!+\! \bpm M(\v p ) \tau_z \!+\! A_1
p_z \tau_x &
A_2 p_-\tau_x \\
A_2 p_+ \tau_x  & M(\v p)\tau_z \!-\! A_1 p_z \tau_x \epm
\label{eq:3D-H} \ea
in the basis spanned by $(\mathrm{Bi}^+_\uparrow,
\mathrm{Se}^-_\uparrow, \mathrm{Bi}^+_\downarrow,
\mathrm{Se}^-_\downarrow)$. Pauli matrices $\bm \tau$ are introduced
and $p_{\pm} = p_x \pm i p_y$ are momentum operators. The upper and
lower indices in the basis set refer to the parity and spin quantum
numbers for the $p_z$ orbitals of Bi or Se atoms, respectively. It
was shown\cite{Bi2Se3-theory} that $\varepsilon(\v p)$ and $M(\v p)
$ depend on the momentum $\v p$ as

\ba \varepsilon (\v p) &=& C+ D_1 p_z^2 + D_2 (p_x^2 + p_y^2 ), \nn
M (\v p) & = & M_0 - B_1 p_z^2 - B_2 (p_x^2 + p_y^2 ) . \ea
Values of the various constants can be found in Refs.
\onlinecite{Bi2Se3-theory,linder,liu,shen,Bi2Se3-theory2}. In our
paper all the material parameters are re-scaled in terms of the one
mass scale $M_0$. Two length parameters emerge as a result, $l_z =
A_1/M_0$ and $l_\perp = A_2/M_0$,  each characterizing the length
scale within the plane and perpendicular to it. With the material
parameters given in Ref. \onlinecite{Bi2Se3-theory} they read
$l_\perp = 14.64${\AA} and $l_z = 7.9$\AA. We use them as the
measure of length in each direction. All equations can be cast in
dimensionless form as well as the two functions $\varepsilon(\v p)$
and $M(\v p)$ which now become (following the parameterization of
Ref. \onlinecite{Bi2Se3-theory})

\ba \varepsilon (\v p)&=& -0.024 + 0.075 p_z^2 + 0.3265 p_\perp^2 ,
\nn M(\v p) &=& 1 - 0.58 p_z^2 -0.94 p_\perp^2 . \ea
Coefficient-by-coefficient, expressions in $\varepsilon(\v p)$ are
smaller than the ones in $M(\v p)$. In this study, we will ignore
$\varepsilon(\v p)$ for calculational simplicity and restore
particle-hole symmetry of the spectrum as a consequence.

The four-dimensional single-particle eigenstates can be constructed
in terms of two, two-dimensional spinors $u$ and $v$. For an
infinite medium one can write the eigenstate as $\psi = e^{i \v k
\cdot \v r} \chi $, where $\chi = \bpm u \\ v \epm$ is a 4-component
constant spinor to be determined by solving

\ba k_+ u &=& \Bigl(E \tau_x + k_z + iM \tau_y \Bigr) v, \nn
 k_-  v &=& \Bigl(E  \tau_x - k_z + iM \tau_y \Bigr) u, \nn
M &=& 1 - \alpha_z k_z^2 - \alpha_\perp k_\perp^2,
\label{eq:3D-eq-dimensionless}\ea
with $E$ as the energy, $k_{\pm} = k_x \pm ik_y$ as the momentum, and
$\alpha_z$ and $\alpha_\perp$ are two material constants. They read
$\alpha_z = 0.58$ and $\alpha_\perp =0.94$ in the parametrization of
Ref. \onlinecite{Bi2Se3-theory}.

\begin{figure}[t]
\includegraphics[scale=0.8]{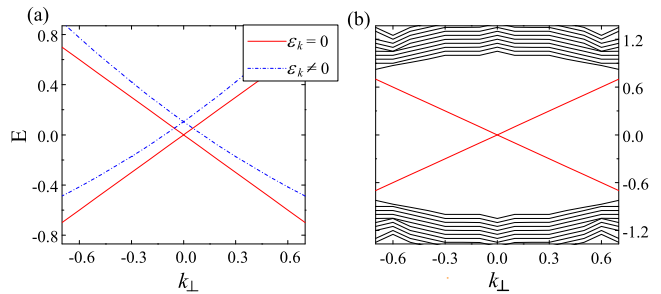}
\caption{(color online) (a) Surface energy spectra without magnetic
field when $\varepsilon_k=0$ (red) and $\varepsilon_k\neq0$ (blue).
(b) Bulk and surface energy dispersions in the absence of magnetic
field and $\varepsilon_k=0$. $L_z/l_z$=3000 was used. }
\label{fig:2D-zero-field-energy-Lz=3000lz}
\end{figure}

As our interest lies in the case of finite thickness $L_z$ for the
$z$-direction the above equation will be deformed as $ik_z
\rightarrow \lambda_z$\cite{linder,liu,shen}. Due to the boundary
conditions at $z=\pm L_z/2$, surface state solutions
appear\cite{linder,liu,shen}. Figure
\ref{fig:2D-zero-field-energy-Lz=3000lz}(a) shows the difference in
the surface energy spectra when the diagonal energy $\varepsilon_k$
is turned on/off. As stressed earlier we will suppress the diagonal
energy $\varepsilon_k$ and work with the particle-hole symmetric
model in the following section where we consider the magnetic field
effect. Figure \ref{fig:2D-zero-field-energy-Lz=3000lz}(b) shows the
surface state energy together with the bulk energy as a function of
the transverse momentum $k_\perp$. The edge state energy dispersion
is precisely linear in $|k_\perp|$ for large $|k_\perp|$ but opens
an exponentially small hybridization gap at $k_\perp =0$. The gap at
the $\Gamma$-point is given by\cite{shen}

\ba\label{eq:2D-zero-field-gap} \Delta  &=&
{8 \alpha\over\beta}\ e^{-\alpha L_z}|\sin(\beta L_z)|, %\nn
%&\simeq&6.96311\ e^{-\alpha L_z}|\sin(\beta L_z)|.
\ea
where $\alpha$, $\beta$ are

\ba\label{eq:2D-zero-field-ab} \alpha={1\over 2\alpha_z},~~
\beta={\sqrt{4\alpha_z-1}\over 2\alpha_z}. \ea
This result will be generalized in the following section to the
nonzero magnetic field, with the revised meaning for the gap as the
energy difference of symmetric and anti-symmetric combinations of
zeroth-Landau levels localized to top and bottom surface layers.

\section{Landau Levels}\label{sec:LL}

Magnetic field $\v H = (0,0,H)$ perpendicular to the slab modifies
the momentum operator $\v p \rightarrow \v p + \v A = (p_x- H y, p_y
, p_z)$ in the Hamiltonian. A pair of canonical operators

\ba \mathcal{A} ={1\over \sqrt{2}} \Bigl({y\!-\!y_0 \over l_H} \!+\!
l_H
\partial_y\Bigr), ~ \mathcal{A}^\dag ={1\over \sqrt{2}}
\Bigl({y\!-\!y_0 \over l_H}  \!-\!l_H \partial_y\Bigr), \ea
are introduced such that $[\mathcal{A} , \mathcal{A} ^\dag]=1$. The
magnetic length (measured in units of $l_\perp$) appears as
$l_H=1/\sqrt{H}$, as well as the guiding center $y_0=l_H^2 k_x$.
Relation to the physical field strength $H_\mathrm{phys}$ in Tesla is

\ba H={l_\perp^2 e H_\mathrm{phys}/ \hbar}\simeq 3.25\times 10^{-3}
H_\mathrm{phys}/[\mathrm{T}]. \ea
Taking $k_+=-(\sqrt{2}/ l_H)\mathcal{A} ^\dag$ and $k_-=-(\sqrt{2}/
l_H)\mathcal{A}$, the eigenvalue equation for a slab with
perpendicular magnetic field becomes

\ba  \mathcal{A}^\dag  u &=& -{l_H\over \sqrt{2}}\Bigl(E \tau_x -i
\lambda_z + i M_{\hat{N}} \tau_y \Bigr) v ,\nn
\mathcal{A} v &=&-{l_H\over \sqrt{2}} \Bigl(E \tau_x + i \lambda_z+
i M_{\hat{N}}\tau_y \Bigr) u,  \label{eq:eq-for-u-and-v}\ea
with several new definitions ($\hat{N}=\mathcal{A}^\dag\mathcal{A}$)

\ba \alpha_{H}={ 2\alpha_\perp \over l_H^2}, ~~ M_{\hat{N}}=
1+\alpha_z \lambda_z^2 -\alpha_{H} \Bigl(\hat{N}+{1\over 2}\Bigr).
\ea
The rest of this section is concerned with the solution of this
equation, together with the boundary conditions at the two
terminations $z = \pm L_z /2$.

The structure of the equation invites for a solution of the form
$u = \phi_{n-1} \bpm a_n \\
b_n \epm$, and $v = \phi_{n} \bpm c_n \\ d_n \epm$, where $\phi_n$
is the $n$-th Landau level (LL) oscillator wave function centered at
$y=y_0$. By substituting the ansatz to Eq. (\ref{eq:eq-for-u-and-v})
we get\cite{comment}

\ba && \sqrt{n} \bpm a_n \\ b_n \epm  = -{l_H\over
\sqrt{2}}\Bigl(E \tau_x \!-\!i \lambda_z \!+\! i M_{n} \tau_y \Bigr) \bpm c_n \\
d_n \epm ,\nn && \sqrt{n} \bpm c_n \\ d_n \epm =-{l_H\over \sqrt{2}}
\Bigl(E \tau_x \!+\! i \lambda_z\!+\! i M_{n-1}\tau_y \Bigr) \bpm a_n
\\ b_n \epm, \nn
&& M_{n}= 1+\alpha_z \lambda_z^2 -\alpha_{H} \Bigl(n+{1\over
2}\Bigr) \label{eq:eq-for-an-bn-cn-dn} . \ea

We can parameterize the spinor solution $u$ and $v$ satisfying Eq.
(\ref{eq:eq-for-an-bn-cn-dn}) in the following form

\ba
u= \phi_{n-1} \cos \varphi \bpm -i\sin {\theta}\\
\cos{\theta}\epm, ~~ v= \phi_{n} \sin \varphi \bpm \cos {\theta} \\
i\sin{\theta}\epm \label{eq:u-and-v-2D-with-field}\ea
with the two complex angles $(\theta,\varphi)$ fixed by

\ba \tan{\theta}={\lambda_z \over E+\mu }, ~~~\tan\varphi = -{{
M_{n-1} \!+\!\mu}\over{\sqrt{2n}/ l_H }}. \label{eq:tangents}\ea
Here $\mu$ means

\ba  \mu \alpha_H =M_n^2-E^2-\lambda_z^2+\alpha_H M_n + {2n \over
l_H^2}. \ea
The eigenvalues are fixed up by the relation $\mu^2 = E^2 +
\lambda_z^2$ which reads when $\mu$ is explicitly written out

\ba \left( M_n^2\!-\!E^2\!-\!\lambda_z^2+\alpha_H M_n \!+\! {2n
\over l_H^2} \right)^2 = \alpha_H^{2} (E^2 \!+\! \lambda_z^2) .
\label{eq:2D-lambdaz-E}\ea
This is the desired characteristic equation for the energy $E$.

Being eighth-power in $\lambda_z$, one can find eight different
$\lambda_z$'s for a given energy.  We call them  $a \lambda_{b}$ as
in the non-magnetic case\cite{linder,liu,shen}, with $a = \pm$ and
$b=1,2,3,4$. There are thus eight independent solutions of the same
energy $E$ for a given LL index $n$ and the guiding center $y_0$,

\ba
\chi_{nab}(y\!-\!y_0) &=& \bpm -ia \sin{\theta_b} \cos\varphi_b \phi_{n-1} \\
\cos {\theta_b }\cos\varphi_b \phi_{n-1}\\ \cos {\theta_b
}\sin\varphi_b \phi_{n}\\ i a\sin{\theta_b} \sin \varphi_b
\phi_{n}\epm . \label{eq:chi-nab}\ea
Taking the linear combination among the eight states gives out the
most general eigenstate before the boundary condition is imposed as

\ba \psi_{nk_x} (x,y-y_0,z) = e^{ik_x x} \sum_{ab} A_{ab} e^{a
\lambda_b z} \chi_{nab} (y-y_0 ). \label{eq:general-form}\ea
To facilitate the further solution, the coefficients $A_{ab}$ can be
classified into symmetric ($A_{ab} = A_b$) and anti-symmetric
($A_{ab} = a A_b$) types. For the symmetric case the boundary
conditions at $z= \pm L_z /2$, $\psi_{nk_x} (x,y-y_0,L_z/2) =
\psi_{nk_x} (x,y-y_0 ,-L_z/2)=0$, can be satisfied if we require
that $A_b$ obey

\ba \sum_{b} A_{b} \sinh ( \lambda_b L_z /2 ) \sin{\theta_b}
\sin\varphi_b = 0,\nn
\sum_{b}  A_{b} \sinh ( \lambda_b L_z /2 ) \sin{\theta_b}
\cos\varphi_b  =0 ,\nn
\sum_{b} A_{b} \cosh ( \lambda_b L_z /2 ) \cos{\theta_b}
\sin\varphi_b = 0, \nn
\sum_{b} A_{b} \cosh ( \lambda_b L_z /2 ) \cos{\theta_b}
\cos\varphi_b = 0. \label{eq:bc-equation}\ea
A nontrivial solution exists provided the characteristic equation of
the above 4$\times$4 matrix is zero. With the aid of Eq.
(\ref{eq:tangents}) this condition can be expressed as
\begin{widetext}

\ba
&~&\Bigl({\lambda_1\lambda_2\tanh{ \lambda_1 L_z \over 2 }\tanh {
\lambda_2 L_z \over 2
}\over(E_n\!+\!\mu_{n,1})(E_n\!+\!\mu_{n,2})}+{\lambda_3\lambda_4\tanh{
\lambda_3 L_z \over 2 }\tanh { \lambda_4 L_z \over 2
}\over(E_n\!+\!\mu_{n,3})(E_n\!+\!\mu_{n,4})}\Bigr)
(\mu_{n,1}\!-\!\mu_{n,2}\!+\alpha_z (\lambda_1^2 - \lambda_2^2 ))
(\mu_{n,3}\!-\!\mu_{n,4}\!+\alpha_z (\lambda_3^2 - \lambda_4^2 ))\nn
&+&\Bigl({\lambda_1\lambda_4\tanh{ \lambda_1 L_z \over 2 }\tanh {
\lambda_4 L_z \over 2
}\over(E_n+\mu_{n,1})(E_n+\mu_{n,4})}+{\lambda_2\lambda_3\tanh{
\lambda_2 L_z \over 2 }\tanh { \lambda_3 L_z \over 2
}\over(E_n\!+\!\mu_{n,2})(E_n\!+\!\mu_{n,3})}\Bigr)
(\mu_{n,1}\!-\!\mu_{n,4}\!+\alpha_z (\lambda_1^2 - \lambda_4^2 ))
(\mu_{n,2}\!-\!\mu_{n,3}\!+\alpha_z (\lambda_2^2 - \lambda_3^2 ))\nn
&=&\Bigl({\lambda_1\lambda_3\tanh{ \lambda_1 L_z \over 2 }\tanh {
\lambda_3 L_z \over 2}
\over(E_n+\mu_{n,1})(E_n\!+\!\mu_{n,3})}+{\lambda_2\lambda_4\tanh{
\lambda_2 L_z \over 2 }\tanh { \lambda_4 L_z \over 2
}\over(E_n\!+\!\mu_{n,2})(E_n\!+\!\mu_{n,4})}\Bigr)
(\mu_{n,1}\!-\!\mu_{n,3}\!+\alpha_z (\lambda_1^2 - \lambda_3^2 ))
(\mu_{n,2}\!-\!\mu_{n,4}\!+\alpha_z (\lambda_2^2 - \lambda_4^2
)),\nn
\label{eq:char-eq-for-LL} \ea
where $M_{n,b}=1+\alpha_z\lambda_b^2-\alpha_H (n+1/2)$, and $
\mu_{n,b} \alpha_H = M_{n,b}^2-E_n^2-\lambda_b^2+\alpha_H M_{n,b} +
2n/l_H^2$.
\end{widetext}

The case of anti-symmetric coefficients $A_{ab} = a A_b$ can be
handled by interchanging $\sinh (\lambda_b L_z /2)$ and
$\cosh(\lambda_b L_z /2)$ in Eq. (\ref{eq:bc-equation}), and
replacing $\tanh(\lambda_bL_z/2)$ by $\coth(\lambda_bL_z/2)$ in Eq.
(\ref{eq:char-eq-for-LL}). Equation (\ref{eq:char-eq-for-LL}) and
its anti-symmetric counterpart can be solved numerically for given
$n$, giving out simultaneously surface and bulk energy solutions in
the presence of the field $H$. When $L_z$ becomes large both
$\tanh(\lambda_b L_z /2)$ and $\coth(\lambda_b L_z /2)$ tend to the
same value and we will have a pair of degenerate states for each
energy, each state being localized either at the top or the bottom
surface and not coupled to the opposite layer.

The zeroth-LL $n=0$ requires a separate treatment. In this case $u$
is identically zero, and $v^T = (c_0, d_0)$ is found from solving

\ba && -{l_H\over
\sqrt{2}}\Bigl(E \tau_x \!-\!i \lambda_z \!+\! i M_{0} \tau_y \Bigr)
\bpm c_0 \\
d_0 \epm =0,\ea
with $M_0=1+\alpha_z\lambda_z^2-\alpha_H /2$. For a given energy
$E_0$, $E_0^2=M_0^2 -\lambda_z^2 $ results in four different
$\lambda_z$'s, $a\lambda_b$ with $a=\pm$ and $b=1,2$. Similar to Eq.
(\ref{eq:chi-nab}) one can assume the spinor solution for $n=0$

\ba \chi_{0ab}(y-y_0) = \phi_0 (y-y_0) \bpm 0 \\ 0 \\ \cos \theta_b \\
i a \sin \theta_b \epm \ea
where $\theta_b$ is given by
\ba \tan \theta_b = {\lambda_b \over E+M_{0,b}},  \ea
and

\ba M_{0,b}&=&1+\alpha_z\lambda_b^2-\alpha_H /2, \nn
\lambda_b&=&{1\over \sqrt{2}\alpha_z}\Bigl(
1-2\alpha_z+\alpha_\perp'\alpha_z \nn &&
-(-1)^b\sqrt{1-4\alpha_z+2\alpha_H \alpha_z+4E^2\alpha_z^2}
  \Bigr)^{{1\over 2}}. \label{eq:2D-n=0-lambdas} \ea
A linear combination

\ba \psi_{0k_x}(y-y_0) = \sum_{ab} A_{ab} e^{a\lambda_b z}
\chi_{0ab}(y-y_0)\ea
can be formed with the boundary conditions at $z=\pm L_z /2$. Again
assuming symmetric ($A_{ab}=A_b$) and anti-symmetric ($A_{ab}=aA_b$)
coefficients separately and denoting the corresponding energies by
$E_0^S$ and $E_0^A$, we have

\ba\label{eq:2D-n=0-sym-BC}
{\lambda_2 \over \lambda_1} { E_0^S\!+\!M_{0,1} \over
E_0^S\!+\!M_{0,2} }&=& {\tanh {\lambda_1L_z\over2}\over \tanh
{\lambda_2L_z\over2}},\ea
and
\ba {\lambda_2 \over \lambda_1} { E_0^A\!+\!M_{0,1} \over
E_0^A\!+\!M_{0,2} }&=& {\tanh {\lambda_2L_z\over2}\over \tanh
{\lambda_1L_z\over2}}.\label{eq:2D-n=0-antisym-BC} \ea
One can easily show from Eqs. (\ref{eq:2D-n=0-sym-BC}) and
(\ref{eq:2D-n=0-antisym-BC}) that $E_0^A=-E_0^S$.

This completes the derivation of the full energy spectra and
eigenstates for a thin-slab geometry of TI model with the
perpendicular magnetic field. In the following section we discuss
several physical results obtained from the analysis of the solution.

\section{Physical Results}
\label{sec:results}

\begin{figure}[h]
\includegraphics[scale=0.8]{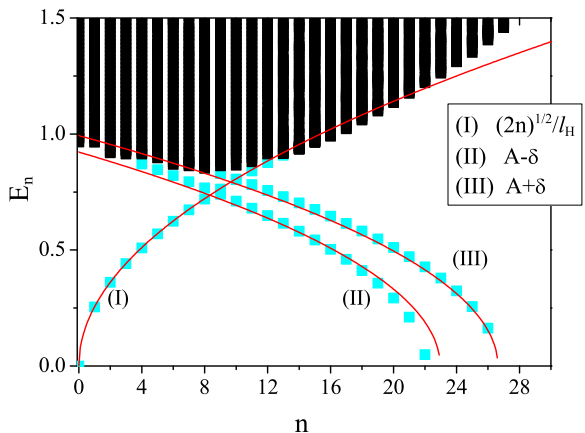}
\caption{(color online)  Landau level energies for
$L_z/l_z=3\times30^3$ and H$_\mathrm{phys}$=10T, showing both
surface(sky blue square) and bulk states(black square). Three
surface branches are labeled (I) through (III) with analytic fits
shown as red solid curves to $\sqrt{2n}/l_H$ (branch I) and
$\sqrt{A\pm \delta - Bn}$ (branches II and III). $A$ and $B$ values
are derived in the Appendix and a small offset $\delta$ is used to
fit the numerical results. When H$_\mathrm{phys}$=10T,
$A=0.918,~\delta=0.068,~B=0.037$, respectively.}
\label{fig:En-n-Lz=21700-H=10T}
\end{figure}

Figure \ref{fig:En-n-Lz=21700-H=10T} shows the dependence of surface
and bulk energies on the LL index $n$, for a sufficiently large
thickness $L_z$. The numerical results remain consistently similar
for $L_z$ larger than about ten times $l_z$. A most surprising aspect
of the numerical analysis is the existence of three distinct branches
of surface-localized states, labeled as (I), (II), and (III) in Fig.
\ref{fig:En-n-Lz=21700-H=10T}.

The behavior of the first surface branch $E^{(1)}_n$ is remarkably
close to the formula:

\ba\label{eq:2D-fit-E}
E^{(s)}_n \simeq\pm\sqrt{2n}/l_H=\pm\sqrt{2nH}.
\ea
This is exactly what is expected of the purely two-dimensional Dirac
Hamiltonian with the Fermi velocity $v_F = 1$ (equal to $M_0 l_\perp
/\hbar=A_2/\hbar\approx 6.2\times10^5$ m/s in physical
units)\cite{Bi2Se3-theory}. Restoring all physical units, the
surface LLs occur at

\ba
E^{(s)}_n  = \pm   {A_2 \over l_{H_\mathrm{phys}}} \sqrt{2n}.
\ea Physical magnetic field $H_\mathrm{phys}=11$T results in the
magnetic length
$l_{H_\mathrm{phys}}=\sqrt{\hbar/eH_\mathrm{phys}}\sim100$\AA~ and
the energy levels $\pm 58\sqrt{n}$ mV. Indeed the spacing in the
$n=0$ and $n=1$ LL peaks were found to be about $40\sim50$
mV\cite{STM-B-thin-film}.

Using the physical magnetic field $H_\mathrm{phys}=10$T we find that
at $n > n_c \approx 12$ the first branch of surface LL begins to
merge with the bulk spectrum (Fig. \ref{fig:En-n-Lz=21700-H=10T}).
Here $n_c$ corresponds to the Landau level index for which the
surface LL begins to touch the bottom of the bulk band. A sharper
criterion to determine $n_c$ can be drawn by keeping track of the
eigenvalues $\lambda_b$ for the surface-bound LLs. With increasing
$n$, one of the four $\lambda_b$'s forming the surface LL eigenstate
has its real part decrease and eventually touch zero at $n=n_c$. This
signals the mixture of an extended state in the wave function just as
the surface LL merges with the bulk continuum. Recently, the number
of surface LLs that can be resolved in the tunneling spectra of
STM\cite{STM-B-thin-film} was shown to be about 12, consistent with
our estimate of $n_c$. For $n>n_c$, the surface branch no longer
exists independently of the bulk LL, but rather seems to form the
bottom of the bulk band as depicted in Fig.
\ref{fig:En-n-Lz=21700-H=10T}.

A recent paper by Liu \textit{et al.} also computed the bulk and
surface LLs based on the model Hamiltonian for Bi$_2$Se$_3$. In their
Fig. 7 it appeared as though the surface LLs can exist well inside
the bulk spectra as an independent branch. We believe this is an
artifact of their calculation not taking care of the boundary
conditions precisely. Once the boundary conditions at $z=\pm L_z /2$
are handled properly, the correct energy profile for $n>n_c$ is the
one in which the surface-localized wave functions are hybridized with
the extended states to form a ``hybrid" state. To confirm this
assertion, we have made a careful analysis of all the $\lambda_b$
values for eigenstates with energies both at the bottom of, and deep
inside the bulk for $n>n_c$. While the details are too tedious to
report here, we can say with certainty that states forming the bulk
LL are typically a linear combination of solutions with real
$\lambda_b$ (localized to surface) and some with purely imaginary
$\lambda_b$ (extended). See Eq. (\ref{eq:general-form}) for a general
definition of the eigenstate. Only for the three surface branches (I)
through (III) is it possible to get all $\lambda_b$'s of the
eigenstate being real and the wave function completely localized.

The existence of extra two surface branches, labeled (II) and (III)
in Fig. \ref{fig:En-n-Lz=21700-H=10T}, is unexpected.  They begin to
appear at $n\approx 4$ and $n\approx8$ respectively for
$H_\mathrm{phys}=10$T. We have confirmed their existence for $L_z
/l_z$ as small as 10 and as large as 3000. Due to the insensitivity
of their features to surface thickness, we can first of all conclude
that the extra surface modes are bound to one particular surface and
not hybridized with the other one. To further confirm that these
branches are genuine, we have carried out an approximate analytic
treatment valid at large LL index $n$ and infinite thickness $L_z$
and indeed found that two extra branches exist. Details of this
analysis are given in the Appendix.

\begin{figure}[h]
\includegraphics[scale=1.0]{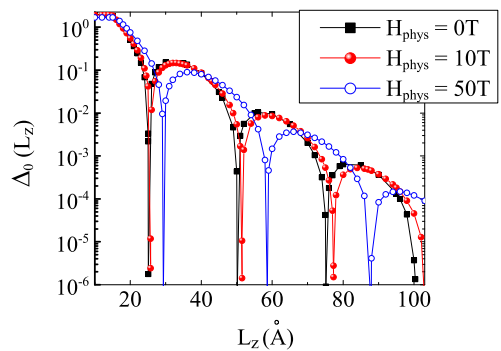}
\caption{(color online) Hybridization gap energies for
$H_\mathrm{phys}=0$T, 10T, and 50T with varying thickness
$L_z$.}\label{fig:gap-n-Lz=21700-H=10T}
\end{figure}

Hybridization effect mixes the two degenerate $n=0$ LLs previously
associated with each surface layer and opens a gap. We have derived
the $n=0$ surface LL energies analytically for symmetric $(E^S_0)$
and anti-symmetric ($E^A_0$) combinations as

\ba
E_0^S&\simeq&-{4\alpha(H)\over\beta(H)}(1\!-\!\alpha_\perp H)\sin
[\beta(H) L_z] e^{-\alpha(H) L_z}, \label{eq:ES-energy}\ea
and $E_0^A = -E_0^S$. The gap is defined as $\Delta_0 = |E_0^S -
E_0^A| =2|E_0^S |$. For practically available field strengths where
$H\ll 1$, $\alpha(H)$ and $\beta(H)$ in Eq. (\ref{eq:ES-energy}) are

\ba\label{eq:2D-alpha-beta}
\alpha(H) \simeq{1\over 2\alpha_z}, ~~ \beta(H)
\simeq{\sqrt{4\alpha_z\!-\!1\!-\!4\alpha_z\alpha_\perp H}\over
2\alpha_z}.\ea
They reduce exactly to $\alpha$ and $\beta$ coefficients obtained in
Eq. (\ref{eq:2D-zero-field-ab}) as $H\rightarrow0$. The gap still
exhibits an oscillatory decay similar to the gap at the
$\Gamma$-point without magnetic field. In Fig.
\ref{fig:gap-n-Lz=21700-H=10T} we compare the energy gaps for
zero-field and for $H_{\mathrm{phys}}=10$T and 50T. The similarity
of their $L_z$-dependence is a strong clue that the origins of the
gaps are the same. Ignoring the small field-induced shift, the gap
can be

\ba \Delta_0 \approx 7M_0 e^{-L_z /[9.2\AA] }\approx 2e^{-L_z
/[9.2\AA] } \mathrm{eV} . \ea
It gives a value $\approx 10$ meV for a seven quintuple-layer thin
film and may well be resolved as two split $n=0$ LLs in a careful
STM spectroscopy study. Currently available thin-film STM study was
done on 50 quintuple-layer sample\cite{STM-B-thin-film}. In Ref.
\onlinecite{liu} it was argued that the oscillation in the sign of
the hybridization gap under zero magnetic field marks the transition
between topologically trivial and non-trivial insulator phases. If
this is so, our calculation seems to reveal that well-defined
Dirac-like LLs exists regardless of the thickness and the sign of
the gap, implying that changes in the topological character of the
thin-film TI will not be revealed by examination of the surface LLs
alone.

\begin{figure}[h]
\includegraphics[scale=0.75]{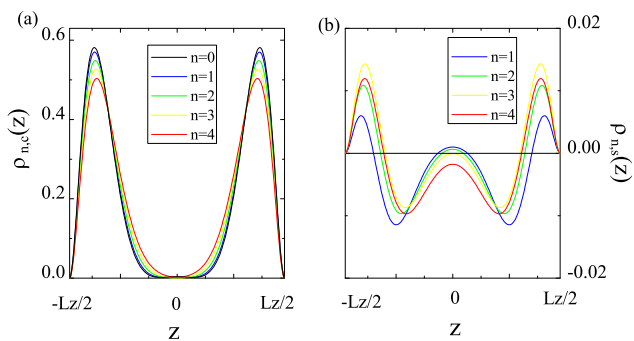}
\caption{(color online) $n$th-LL wave function density
$\rho_{n,c}(z)$ and the spin density $\rho_{n,s}(z)$for $L_z=60$\AA.
} \label{fig:spin-density}
\end{figure}

The charge $(c)$ and spin $(s)$ densities of each $n$-th surface LL
wave function can be defined as

\ba \rho_{n,s(c)} (z) &=& {\int dx dy ~ \psi^\dag_{n} \Gamma_{s(c)}
\psi_{n} \over \int dx dy dz ~ \psi^\dag_{n} \psi_{n} }, \nn
\Gamma_{s} &=& \mathrm{diag}(1,1, -1,-1), \nn
\Gamma_{c} &=& \mathrm{diag}(1,1, 1,1).\ea
Figure \ref{fig:spin-density} shows results for a few surface LLs
with small LL index $n$. All surface LLs are localized to within one
$l_z$ of the termination, or within about one quintuple layer. As one
can see from Fig. \ref{fig:spin-density}(b), the zeroth-LL is
completely spin-polarized, $\int \rho_{0,s}(z) dz =-1$, while other
higher surface LLs are nearly spin-quenched, $\int \rho_{n>0,s}(z) dz
\approx 0$. The zeroth-LL has only the lower two elements of the
four-component spinor $\chi$ take nonzero values, which refer to the
amplitudes for Bi and Se states of spin-$\downarrow$ (See text
following Eq. \ref{eq:3D-H}). There are two $n=0$ LL in the solution,
and both of them are fully spin-$\downarrow$-polarized. The origin of
the spin polarization is the analogue of the sublattice polarization
of the $n=0$ LL in graphene\cite{castro-neto}. The difference is that
the two valley $n=0$ Landau levels occupy the opposite sublattices,
so that the overall sublattice symmetry is restored.

Here, by contrast, both top and bottom surface LLs give the
\textit{same spin polarization}. The reason is that for the top
surface Dirac states the magnetic field is pointing \textit{out of
the bulk} but the bottom surface states experience the field
pointing \textit{into the bulk}, so that effectively the sense of
the field direction is also reversed between the two surface layers.
By reversing the field direction from $+\hat{z}$ to $-\hat{z}$ one
will generate $n=0$ of spin-$\uparrow$ polarization. As a result a
thin slab of TI subject to quantizing magnetic field creates two
$n=0$ LLs which are completely spin polarized. Such spin-polarized
surface layers are detectable by Faraday or Kerr rotation
experiments\cite{Bi2Se3-theory}.

\section{Conclusion}
\label{sec:conclusion}

We showed how to derive the Landau level solution for a slab geometry
of the topological insulator based on the four-band
model\cite{Bi2Se3-theory,Bi2Se3-theory2}. Previoius approaches were
to first project the zero-field bulk Hamiltonian to the surface, then
using the Peierls substitution to address the magnetic field
effect\cite{SQShen-LL,Bi2Se3-theory2}. Our strategy by contrast is to
introduce the Peierls substitution directly into the bulk Hamiltonian
and use the boundary conditions appropriate for a slab geometry. The
obtained surface Landau level energies are in good accord with those
obtained from the surface Dirac Hamiltonian, and we conclude that
surface projection and the Peierls substitution can be implemented in
any order with the same physical spectrum.

A dramatic departure of the present Dirac LL problem with an
analogous one posed by the graphene system\cite{castro-neto} is that
the surface LLs are eventually bounded by the bulk spectra, and one
has to face the issue what will happen to the surface LLs  as they
begin to merge with the bulk continuum. We addressed such a question
numerically and analytically in this paper, with a prediction for the
existence of new surface-bound LLs appearing at higher-LL indices.
Detection of the predicted new surface modes presents an interesting
challenge for the future surface-sensitive measurements on TI
materials.

\acknowledgments  H. J. H. is supported by Mid-career Researcher
Program through NRF grant funded by the MEST (No.
R01-2008-000-20586-0).

\appendix
\section{Analysis of New Surface Modes}
After some trial and error, we find that the following ansatz
describe the numerically found surface modes (2) and (3) with good
accuracy.

\ba\label{eq:ansatz-lambda}\lambda_b^2= {\alpha_H \over \alpha_z }
(n-m_b \sqrt{n}  ). \ea
Here $m_b$ is a constant, to be determined later. This gives for
$M_n$,

\ba M_n = 1 - \alpha_H  m_b \sqrt{n} - {1\over 2} \alpha_H .
\label{eq:ansatz-for-Mn}\ea
We can also make an ansatz for the surface energy mode of the form

\ba E_n^2 = A-B n, \label{eq:ansatz-for-En} \ea
with two undetermined positive coefficients $A$ and $B$. Inserting
Eqs. (\ref{eq:ansatz-lambda}) through (\ref{eq:ansatz-for-En}) into
Eq. (\ref{eq:2D-lambdaz-E}) gives

\begin{widetext}

\ba \left[ \left( {\alpha^2_H }m_b^2 + B -{\alpha_H \over \alpha_z }
+ {2\over l_H^2} \right) n + \left({1\over \alpha_z}
-2\right)\alpha_H m_b \sqrt{n} +\cdots \right]^2 = \alpha_H^2
\left({\alpha_H \over \alpha_z } -B \right) n + \cdots . \ea
\end{widetext}
The terms in $\cdots$ have subleading order in $n$ than the ones
shown. Assuming a sufficiently large $n$ we require that the two
sides of the equation cancel out at each order in $n$. From the
equality of $n^2$, $n$, and $\sqrt{n}$-order terms we obtain the
three following equations:

\ba (\alpha_H m_b )^2 &= &{\alpha_H \over \alpha_z } - {2\over l_H^2}
- B ,\nn
\left({1\over \alpha_z} -2\right)^2 (\alpha_H  m_b)^2 &=& \alpha_H^2
\left({\alpha_H \over \alpha_z } -B \right),\nn
2(1-{\alpha_H^2\over4}-A)({1\over\alpha_z}-2)\alpha_H^2m_b&=&
-{\alpha_H\over\alpha_z}\alpha_H^2m_b .\label{eq:B-and-mb}\ea
Upon solving them we obtain

\ba m_b^2 &=&{2/l_H^2\over(2-1/\alpha_z)^2-\alpha_H^2} ,\nn
B&=&{2\over l_H^2}-
{\alpha_H\over\alpha_z}-{2\alpha_H^2/l_H^2\over(2-1/\alpha_z)^2-\alpha_H^2},
\nn
A&=&1-{\alpha_H^2\over4}+{2\alpha_H^2/\alpha_z\over1/\alpha_z-2}.\ea
Further consideration of sub-leading corrections finally yield a
splitting of $A$ into two branches responsible for (II) and (III) in
Fig. \ref{fig:En-n-Lz=21700-H=10T}. Rather than going into the
complicated sub-leading order analysis, we can simply split $A$ into
two branches by writing $A\pm \delta$ with $\delta$ chosen to fit the
two branches in Fig. \ref{fig:En-n-Lz=21700-H=10T} while $A$ itself
is completely determined from the parameters such as $\alpha_z$ and
$\alpha_\perp$.  It is shown that both branches (II) and (III) match
quite well the ansatz for energy, Eq. (\ref{eq:ansatz-for-En}).

We can also discuss the stability of the new surface branches by
recalling the numerical value $\alpha_z = 0.58$, and $\alpha_H =
2\alpha_\perp/l_H^2 = 1.84/l_H^2$. It follows that positive
$(\alpha_H m_b)^2$ is possible if $2-(0.58)^{-1}-1.84/l_H^2
=0.276-1.84/l_H^2 > 0$, or if $l_H^2 > 6.66$. Returning to physical
length scales, this implies the magnetic length greater than
$\sim$38\AA, or the magnetic field strength less than 46T. We then
expect that surface modes (II) and (III) should co-exist with the
more familiar mode (I) inside the bulk gap for typical laboratory
magnetic field ranges.


\begin{thebibliography}{99}
\bibitem{comment} It follows immediately that an eigenstate with energy $-E$ is
obtained from the state with energy $E$ in Eq. (9)  by the operation
$\tau_y$.
\end{thebibliography}
\end{document}